\documentclass[aps,onecolumn,pra]{revtex4}
\usepackage{amsmath,amssymb}
\usepackage[dvips]{epsfig}
\usepackage{graphicx}

{\hspace*{\fill}$\rule{.3\baselineskip}{.35\baselineskip}$\end{trivlist}}

\newcommand{\R}{\mathbb{R}}

\newcommand{\Z}{\mathbb{Z}}
\newcommand{\N}{\mathbb{N}}

\renewcommand{\geq}{\geqslant}

\renewcommand{\phi}{\varphi}
\newcommand{\be}{\begin{eqnarray}}
\newcommand{\ee}{\end{eqnarray}}

\newcommand{\eps}{\varepsilon}

\begin{document}

\title{\bf Distribution of eigenfrequencies for oscillations of the ground state in the
Thomas--Fermi limit}

\author{P.G. Kevrekidis$^1$ and D.E. Pelinovsky$^2$\\
{\small $^{1}$ Department of Mathematics and Statistics, University
of Massachusetts, Amherst, MA 01003} \\
{\small $^{2}$ Department of Mathematics and Statistics, McMaster
University, Hamilton, Ontario, Canada, L8S 4K1}  }

\date{\today}

\begin{abstract}
In this work, we present a systematic derivation of
the distribution of eigenfrequencies for oscillations of the ground state of a
repulsive Bose-Einstein condensate in the semi-classical (Thomas-Fermi)
limit. Our calculations are performed in 1-, 2- and 3-dimensional
settings. Connections with the earlier work of Stringari,
with numerical computations, and with theoretical expectations for
invariant frequencies based on symmetry principles are also given.
\end{abstract}

\maketitle

\section{Introduction}

Bose-Einstein condensation (BEC) is one of the most exciting
achievements within the physics community in the last two decades.
Its experimental realization in 1995, by
two experimental groups using vapors of Rb \cite{anderson} and
Na \cite{davis} marked the formation of a new state of matter consisting of
a cloud of atoms within the same quantum state, creating a giant
matter wave. However, in addition to its impact on the physical
side, this development had a significant influence on mathematical
studies of such Bose-Einstein condensates (BECs) \cite{book1,book2,review,rcg:BEC_BOOK,rcg:65}.
In considering typical BEC experiments and in exploring 
the unprecedented control
of the condensates through magnetic and optical ``knobs'', a mean-field theory
is applied to reduce the quantum many-atom description to a scalar nonlinear
Gross-Pitaevskii equation (GPE). This is a variant of the famous nonlinear Schr\"odinger (NLS)
equation \cite{sulem,ablowitz} of the form:
\begin{equation}
i \hbar \frac{\partial\Psi}{\partial t}=-\frac{{\hbar}^2}{2 m} \nabla^2 \Psi
+ g |\Psi|^2 \Psi + V_{{\rm ext}} ({\bf r}) \Psi,
\label{peq1}
\end{equation}
where $\Psi=\Psi({\bf r},t)$ is the BEC wavefunction
(the atomic density is proportional to $|\Psi({\bf r},t)|^2$),
$\nabla^2$ is the Laplacian in ${\bf r} = (x,y,z)$, $m$ is the atomic mass,
the prefactor $g$ is proportional to the atomic scattering
length (e.g. $g>0$ for Rb and Na, while $g<0$ for Li atoms), and $V_{\rm ext}({\bf r})$
is the external potential for magnetic or optical traps.

The NLS equation is a well-established model in applications
in optical and plasma physics as well as in fluid mechanics,
where it emerges out of entirely different physical considerations
\cite{sulem,ablowitz}.
In particular, for instance,
in optics, it emerges due to the so-called Kerr effect, whereby
the material refractive index depends linearly on the intensity
of incident light. The widespread use of the NLS equation stems from the fact
that it describes, to the lowest order, the nonlinear dynamics of envelope waves.

One of the particularly desirable features of GPE is that
the external potential $V_{\rm ext}({\bf r})$ can assume a multiplicity
of forms, based on the type of trapping used to confine the atoms.
Arguably, however, the most typical magnetic trapping imposes a parabolic
potential \cite{review,kevfra}
\begin{equation}
V_{{\rm ext}}=\frac{m}{2} ( \omega_x^2 x^2 + \omega_y^2 y^2 + \omega_z^2 z^2),
\label{peq2}
\end{equation}
where, in general, the trap frequencies $\omega_{x,y,z}$
are different. In what follows, however, for simplicity, we will
restrict our consideration to the isotropic case of equal frequencies
$\omega_x = \omega_y = \omega_z \equiv \omega$ along the different directions.

If the wave function is decomposed according to
$$
\Psi({\bf r},t) = e^{-i \mu t/\hbar} U({\bf r}),
$$
then $U({\bf r})$ solves the stationary GPE with the chemical potential $\mu$.
One of the most extensively discussed limits in the case of self-repulsive
nonlinearity $g > 0$ and in the presence
of the parabolic potential (\ref{peq2}), is the limit $\mu \to \infty$. This limit
is referred to as the Thomas--Fermi limit.

If the kinetic (Laplacian) term is neglected, the approximate ground state
solution is obtained in the form
$$
U({\bf r}) = \left( \max[g^{-1} (\mu-V_{{\rm ext}}({\bf r})),0] \right)^{1/2}.
$$
For this limit, the seminal work of Stringari \cite{stringari}
suggested a computation of the corresponding eigenfrequencies of
oscillations of perturbations around the ground state of the system, using a
hydrodynamic approach. This approach has become popular in the physics literature for more
complicated problems involving anisotropic traps \cite{FCSG} and
dipole--dipole interactions \cite{EGD}.

The aim of the present work is to derive these eigenfrequencies
systematically not only in the 3-dimensional context,
but also in the 2-dimensional and 1-dimensional cases.
We will relate these eigenfrequencies to the eigenvalues discussed
in the recent work \cite{GalPel1} in the 1-dimensional setting.
The relevant eigenfrequencies of the perturbations around the ground state
are then directly compared with numerical computations in the 2-dimensional case.

In the numerical computations, the eigenfrequencies are obtained systematically
as a function of the chemical potential $\mu$ starting from the low-amplitude limit
(when the ground state is approximately that of the parabolic
potential) all the way to the large-chemical potential. Earlier,
these eigenfrequencies were approximated numerically near the low-amplitude
limit by Zezyulin {\em et al.} \cite{ZAKP} in the 1-dimensional case and
by Zezyulin \cite{zezyulin} in the 2-dimensional case.
Our numerical computations also allow us to identify eigenfrequencies
that remain invariant under changes in $\mu$ and to connect them
to underlying symmetries of the GPE.

Our presentation will be structured as follows. In section 2, we
present the mathematical setup of the problem.
In section 3, we compute its
corresponding linearization eigenvalues (around the Thomas-Fermi ground
state) in the 1-, 2- and 3-dimensional settings. In section 4,
we compare these results to direct numerical computations in the
2-dimensional case. Note that the 1-dimensional case was considered in some
detail in our earlier work \cite{PeliKev}. Lastly, a brief summary of our findings
and some interesting directions
for future study are offered in section 5.

\section{Mathematical Setup}

Using rescaling of variables, one can normalize the GPE (\ref{peq1}) into two equivalent forms. One
form corresponds to the semi-classical limit and it arises if $\hbar = \eps$, $m = \frac{1}{2}$,
$g = 1$, $\omega = 2$, and $\mu = 1$, or equivalently, in the form
\begin{equation}
\label{GP} i \eps u_t + \eps^2 \nabla^2 u + (1 - |x|^2 - |u|^2) u = 0,
\end{equation}
where $u(x,t) : \mathbb{R}^d \times \mathbb{R} \to \mathbb{C}$ is a wave function,
$\nabla^2 = \partial_{x_1}^2 + ... + \partial_{x_d}^2$ is the Laplacian operator
in $d$ spatial dimensions, and $\eps$ is a small parameter.
On the other hand, if
\begin{equation}
v = \mu^{1/2} u, \quad \xi = (2 \mu)^{1/2} x, \quad \tau = 2 t,
\label{translation}
\end{equation}
then equation (\ref{GP}) can be translated to the form
\begin{equation}
\label{GPphys}
i v_{\tau}  = - \frac{1}{2} \nabla_{\xi}^2 v + \frac{1}{2} |\xi|^2 v +
|v|^2 v  - \mu v,
\end{equation}
that corresponds to the GPE (\ref{peq1}) with $\hbar = 1$, $m = 1$, $g = 1$, $\omega = 1$, and
$\mu = \frac{1}{2\eps}$. The semi-classical limit $\eps \to 0$ corresponds to the
Thomas--Fermi limit $\mu \rightarrow \infty$.

Let $\eta_{\eps}$ be a real positive solution of the stationary problem
\begin{equation}
\label{stationaryGP} \eps^2 \nabla^2 \eta_\eps + (1- |x|^2 -
\eta_\eps^2) \eta_{\eps} =0, \quad x \in \mathbb{R}^d.
\end{equation}
According to Gallo \& Pelinovsky \cite{GalPel2}, for any sufficiently small $\eps > 0$
there exists a smooth radially symmetric solution $\eta_{\eps} \in {\cal
C}^{\infty}(\mathbb{R}^d)$ that decays to zero as $|x| \to \infty$
faster than any exponential function. This solution
converges pointwise as $\eps \to 0$ to the compact Thomas--Fermi
cloud
\begin{equation}
\label{Thomas-Fermi}
\eta_0 := \lim_{\eps \to 0} \eta_{\eps} =
\left\{ \begin{array}{cl} (1 - |x|^2)^{1/2}, \;\; & \mbox{for} \;\; |x| < 1, \\
0, \;\; & \mbox{for} \;\; |x| > 1. \end{array} \right.
\end{equation}
The solution $\eta_{\eps}$ with the properties above is generally referred to as
the {\em ground state} of the Gross--Pitaevskii equation (\ref{GP}).

The spectral stability problem (often referred to as the Bogolyubov-de Gennes
problem in the context of BECs) for the ground state
$\eta_{\eps}$ is written as
the eigenvalue problem
\begin{equation}
\label{LL}
L_+ u = -\lambda \eps w, \quad L_- w = \lambda \eps u,
\end{equation}
associated with the two Schr\"{o}dinger operators
$$
\left\{ \begin{array}{l} L_+ = -\eps^2 \nabla^2 + |x|^2 + 3 \eta_{\eps}^2 - 1,
\\
L_- = -\eps^2 \nabla^2 + |x|^2 + \eta_{\eps}^2 - 1. \end{array} \right.
$$

A naive approximation of the eigenvalues $\lambda$ of the spectral stability problem (\ref{LL})
arises if we replace $\eta_{\eps}^2$ by $\eta_0^2$. Because $L_+$ is invertible,
the eigenvalue problem can then be written in the form
\begin{equation}
\label{LL-1}
\left( -\eps^2 \nabla^2 + |x|^2 + \eta_0^2 - 1 \right) w = \gamma \eps^2
\left( -\eps^2 \nabla^2 + |x|^2 + 3 \eta_0^2 - 1 \right)^{-1} w, \quad x \in \R^d,
\end{equation}
where $\gamma = -\lambda^2$. The formal limit $\eps \to 0$ gives a restricted problem in the unit ball
\begin{equation}
\label{LL-1-red}
\mbox{LI} : \quad -2(1 - |x|^2) \nabla^2 w = \gamma w, \quad x \in B_0 = \{ x \in \R^d : \;\; |x| < 1 \},
\end{equation}
subject to the Dirichlet boundary condition on the sphere $|x| = 1$. Convergence
of eigenvalues of (\ref{LL-1}) to eigenvalues of the limiting problem (\ref{LL-1-red})
was rigorously justified by Gallo \& Pelinovsky \cite{GalPel1} in one spatial dimension $d = 1$.

Because $L_+$ is invertible for any small $\eps > 0$, the original eigenvalue problem (\ref{LL})
can also be written in the form
\begin{equation}
\label{LL-2}
\left( -\eps^2 \nabla^2 + |x|^2 + \eta_{\eps}^2 - 1 \right) w = \gamma \eps^2
\left( -\eps^2 \nabla^2 + |x|^2 + 3 \eta_{\eps}^2 - 1 \right)^{-1} w, \quad x \in \R^d.
\end{equation}
Because
$$
|x|^2 + \eta_{\eps}^2 - 1 = \frac{\eps^2 \nabla^2 \eta_{\eps}}{\eta_{\eps}},
$$
the formal limit $\eps \to 0$ gives now a different problem in the unit ball
\begin{equation}
\label{LL-2-red}
\mbox{LII} : \quad - 2(1 - |x|^2) \left( \nabla^2 w - \frac{\nabla^2 \eta_0}{\eta_0} w \right) = \gamma w, \quad x \in B_0,
\end{equation}
subject to Dirichlet boundary conditions on the sphere $|x| = 1$. Justification of convergence of
eigenvalues of (\ref{LL-2}) to eigenvalues of the limiting problem (\ref{LL-2-red})
is still an open problem in analysis.

The limiting eigenvalue problem (\ref{LL-1-red}) can be written in the vector form 
\begin{equation}
\label{LL-1-red-vect}
\mbox{LI} : \quad -2 \nabla (1 - |x|^2) \nabla {\bf v} = \gamma {\bf v}, \quad x \in B_0,
\end{equation}
where ${\bf v} = \nabla w \in \mathbb{R}^d$. On the other hand, the limiting eigenvalue 
problem (\ref{LL-2-red}) can be rewritten in the equivalent scalar form
\begin{equation}
\label{LL-3-red}
\mbox{LII} : \quad  -2 \nabla (1 - |x|^2) \nabla v = \gamma v, \quad x \in B_0,
\end{equation}
where $v = \frac{w}{\eta_0}$ and $\eta_0$ is given by (\ref{Thomas-Fermi}). It was exactly the representation 
(\ref{LL-3-red}) of the limiting eigenvalue problem $LII$, which was derived by Stringari \cite{stringari} 
from the hydrodynamical formulation of the Gross--Pitaevskii equation (\ref{GP}) in three dimensions $d = 3$.

Comparison of the two representations (\ref{LL-1-red-vect}) and (\ref{LL-3-red}) implies 
that the two limiting eigenvalue problems have {\em identical} nonzero eigenvalues
in the space of one dimension $d = 1$ but may have {\em different} nonzero eigenvalues for $d \geq 2$.
We will show in the next section that it is exactly the case. 
We will illustrate numerically for $d = 2$ that the eigenvalues of the second limiting 
problem (\ref{LL-3-red}) are detected in the limit $\eps \to 0$ from 
the eigenvalues of the original problem (\ref{LL}).

\section{Eigenvalues of the limiting problems}

{\bf Case $d = 1$:} Both representations (\ref{LL-1-red-vect}) and (\ref{LL-3-red}) 
of the limiting eigenvalue problems $LI$ and $LII$ reduce to the Legendre equation
\begin{equation}
\label{Legendre}
(x^2 - 1) v''(x) + 2 x v'(x) = \frac{1}{2} \gamma v(x), \quad x \in (-1,1).
\end{equation}

For $LI$ given by (\ref{LL-1-red}), the correspondence of eigenfunctions is $v(x) = w'(x)$. 
The only nonsingular solutions of this equation at the regular singular points $x = \pm 1$
are Legendre polynomials $v(x) \in \{ P_n(x) \}_{n \geq 0}$, which correspond to eigenvalues
$\gamma \in \{ 2 n(n+1) \}_{n \geq 0}$. The zero eigenvalue must be excluded from
the set since it corresponds to $w(x) = x$, which violates Dirichlet boundary conditions at
$x = \pm 1$ for $w(x)$. On the other hand, all nonzero eigenvalues are present because the
corresponding eigenfunction $w(x) = C_{n+1}^{-1/2}(x)$ constructed from $v(x) = P_n(x)$
thanks to identities 8.936, 8.938, and 8.939 in \cite{Grad}
$$
\frac{d}{dx} C_{n+1}^{-1/2}(x) = C_n^{1/2}(x) = P_n(x),
$$
also satisfies the Dirichlet boundary conditions thanks to the identity
$$
C_{n+1}^{-1/2}(x) = \frac{x^2-1}{n(n+1)} \frac{d^2}{dx^2} C_{n+1}^{-1/2}(x).
$$

For $LII$ given by (\ref{LL-2-red}), the correspondence of eigenfunctions is 
$v(x) = \frac{w(x)}{\sqrt{1-x^2}}$. The zero eigenvalue should now be included for $n = 0$, 
since the eigenfunction $w(x) = \sqrt{1-x^2}$ corresponds to the ground state $\eta_{\eps}$ 
in the limit $\eps \to 0$, which is known to be the eigenfunction of operator $L_-$. 
All nonzero eigenvalues are the same
as for the limiting problem (\ref{LL-1-red}) but the eigenfunctions are now different. For
the eigenvalue $\gamma = 2n(n+1)$, the eigenfunction is $w(x) = \sqrt{1 - x^2} P_n(x)$. 
It should be noted here that the obtained eigenvalue distribution was numerically examined in \cite{PeliKev}
and was found to be in very good agreement with the true eigenvalues of system (\ref{LL}). 

\vspace{0.5cm}

{\bf Case $d = 2$:} We consider the limiting eigenvalue problem $LII$ in the form (\ref{LL-3-red}) and use
the polar coordinates
$$
\left\{ \begin{array}{l} x = r \cos(\theta), \\ y = r \sin(\theta), \end{array} \right.
\quad r \geq 0, \;\; \theta \in [0,2\pi].
$$
After the separation of variables $v(r,\theta) = V(r) e^{i m \theta}$ for $m \in \Z$, we
obtain an infinite set of eigenvalue problems for amplitudes of cylindrical harmonics
\begin{equation}
\label{LL-2-red-radial}
-(1 - r^2) \left( V''(r) + \frac{1}{r} V'(r) - \frac{m^2}{r^2} V(r) \right) + 2 r V'(r)
= \frac{1}{2} \gamma V(r), \quad r \in (0,1).
\end{equation}
Let $m \geq 0$. We are looking for solutions of equation (\ref{LL-2-red-radial})
which behave like $V(r) \sim r^m$ as $r \to 0$. Let us transform (\ref{LL-2-red-radial})
to a hypergeometric equation with the substitution $V(r) = r^m F(z)$, $z = r^2$. Direct computations
show that $F(z)$ solves
$$
z(1-z) F''(z) + (1 + m - (2+m)z) F'(z) + \left( \frac{1}{8} \gamma - \frac{1}{2} m \right) F(z) = 0, \quad z \in (0,1).
$$
A nonsingular solution at $z = 0$ is the hypergeometric function $F(z) = {\cal F}(a,b,c;z)$ where
$$
c = 1 + m, \quad a + b = 1 + m, \quad ab =  \frac{m}{2} - \frac{\gamma}{8}.
$$
Because $a + b - c = 0$, the hypergeometric function is singular at $z = 1$ unless it
becomes a polynomial for $a = -k$ with an integer $k \geq 0$.
The eigenvalues of the limiting problem (\ref{LL-3-red}) are then given by
\begin{equation}
\label{set-of-eigenvalues-1}
\gamma \in \{ \gamma_{m,k}^{(1)} \}_{m \geq 0, k \geq 0}, \quad
\gamma_{m,k}^{(1)} = 4( m + 2k(1+m) + 2k^2).
\end{equation}

Let us now consider the limiting eigenvalue problem $LI$ in the form (\ref{LL-1-red}) and use the same polar coordinates. The corresponding eigenvalue problem is
\begin{equation}
\label{LL-1-red-polar}
-2(1 - r^2) \left( \frac{\partial^2 w}{\partial r^2} + \frac{1}{r} \frac{\partial w}{\partial r}
+ \frac{1}{r^2} \frac{\partial^2 w}{\partial \theta^2} \right) = \gamma w,
\quad r \in (0,1), \;\; \theta \in [0,2\pi].
\end{equation}
Let $w(r,\theta) = V(r) e^{i m \theta}$ for $m \in \Z$ and obtain an infinite
set of eigenvalue problems for cylindrical harmonics
\begin{equation}
\label{LL-1-red-radial}
-(1 - r^2) \left( V''(r) + \frac{1}{r} V'(r) - \frac{m^2}{r^2} V(r) \right)
= \frac{1}{2} \gamma V(r), \quad r \in (0,1).
\end{equation}
Let $m \geq 0$. Equation (\ref{LL-1-red-radial}) can also be transformed
to a hypergeometric equation after the substitution $V(r) = r^m F(z)$, $z = r^2$. Direct computations
show that $F(z)$ solves
$$
z(1-z) F''(z) + (1 + m)(1-z) F'(z) + \frac{1}{8} \gamma F(z) = 0, \quad z \in (0,1).
$$
A nonsingular solution at $z = 0$ is the hypergeometric function $F(z) = {\cal F}(a,b,c;z)$ where
$$
c = 1 + m, \quad a + b = m, \quad ab =  - \frac{\gamma}{8}.
$$
Because $a + b - c = -1$, the hypergeometric function is bounded at $z = 1$. However,
we need $F(1) = 0$ to satisfy the Dirichlet boundary conditions for $w(r,\theta)$
at $r = 1$. From this condition,
$F(z)$ has to be a polynomial (or $F'(z)$ and $F''(z)$ are singular at $z = 1$ with $F(1) \neq 0$).
The polynomial arises for $a = -k$ with an integer $k \geq 0$.
The eigenvalues of the limiting problem (\ref{LL-1-red}) are then given by
\begin{equation}
\label{set-of-eigenvalues-2}
\gamma \in \{ \gamma_{m,k}^{(2)} \}_{m \geq 0, k \geq 0}, \quad
\gamma_{m,k}^{(2)} = 8k(m + k).
\end{equation}

Comparison of (\ref{set-of-eigenvalues-1}) and (\ref{set-of-eigenvalues-2}) show
that $\gamma_{0,k}^{(1)} = \gamma_{1,k}^{(2)}$ for all $k \geq 0$, but the sets
$\{ \gamma_{m,k}^{(1)} \}_{m \geq 0, k \geq 0}$ and $\{ \gamma_{m,k}^{(2)} \}_{m \geq 0, k \geq 0}$
are different. For instance, $\gamma_{1,0}^{(1)} = 4$ is not present in the set
$\{ \gamma_{m,k}^{(2)} \}_{m \geq 0, k \geq 0}$.

\vspace{0.5cm}

{\bf Case $d = 3$:} We consider the limiting eigenvalue problem $LII$ in the form (\ref{LL-3-red}) and use
the spherical coordinates
$$
\left\{ \begin{array}{l} x = r \cos(\theta) \cos(\varphi), \\ y = r \sin(\theta) \cos(\varphi), \\
z = r \sin(\varphi), \end{array} \right.
\quad r \geq 0, \;\; \theta \in [0,2\pi], \;\; \varphi \in [0,\pi].
$$
After the separation of variables $v(r,\theta,\varphi) = V(r) Y_{l,m}(\theta,\varphi)$ for $m \in \Z$
and $l \in \N$, where $Y_{l,m}(\theta,\varphi)$ are spherical harmonics, we
obtain an infinite set of eigenvalue problems for amplitudes of the spherical harmonics:
\begin{equation}
\label{LL-3-red-radial}
-(1 - r^2) \left( V''(r) + \frac{2}{r} V'(r) - \frac{l(l+1)}{r^2} V(r) \right) + 2 r V'(r)
= \frac{1}{2} \gamma V(r), \quad r \in (0,1).
\end{equation}
Using a similar reduction $V(r) = r^l F(z)$, $z = r^2$ to the hypergeometric equation, we
obtain the eigenvalues of the limiting problem (\ref{LL-3-red}) in the form
\begin{equation}
\label{set-of-eigenvalues-1-3d}
\gamma \in \{ \gamma_{l,k}^{(1)} \}_{l \geq 0, k \geq 0}, \quad
\gamma_{l,k}^{(1)} = 4( l + 3 k + 2kl + 2k^2).
\end{equation}
This distribution was obtained by Stringari \cite{stringari} from the balance of 
the leading powers in polynomial solutions of (\ref{LL-2-red-radial}).

Using the same algorithm, the eigenvalues of the limiting problem $LI$ are found in the form
\begin{equation}
\label{set-of-eigenvalues-2-3d}
\gamma \in \{ \gamma_{l,k}^{(2)} \}_{l \geq 0, k \geq 0}, \quad
\gamma_{l,k}^{(2)} = 4 k (1 + 2l + 2k).
\end{equation}
This distribution is different from (\ref{set-of-eigenvalues-1-3d}). In particular, it does not include
eigenvalue $\gamma_{1,0}^{(1)} = 4$.

\section{Numerical results}

\begin{figure}
\begin{center}
\includegraphics[width=0.9\textwidth]{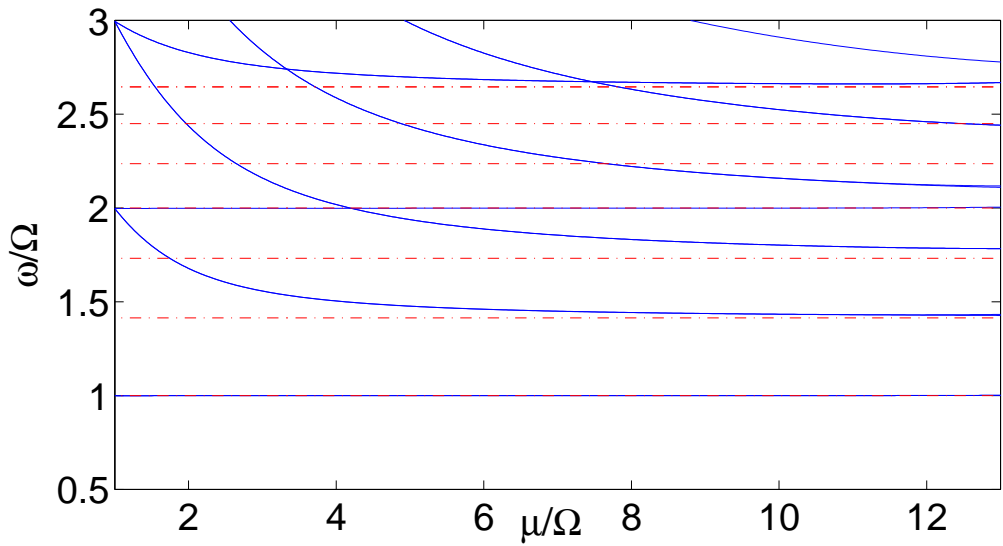}
\end{center}
\caption{Eigenfrequencies of the two-dimensional spectral problem (\ref{LL})
(solid blue lines) and
the limiting eigenvalues of the reduced problem (\ref{LL-2-red}) (dash-dotted
red lines).
Notice that for practical numerical reasons the computation was performed
for a parabolic trap of frequency $\Omega=0.3$ and the corresponding
eigenfrequencies and chemical potentials were appropriately rescaled
by $\Omega$.}
\label{fig1}
\end{figure}

As indicated above in the one-dimensional case, 
good agreement was observed between
the predicted Thomas-Fermi limit spectrum and the numerical computations
of \cite{PeliKev}; for this reason, we now turn our attenion to the 
two-dimensional case.
Eigenvalues of the original spectral problem (\ref{LL}) for $d = 2$ are computed numerically
and shown on Figure \ref{fig1} (solid lines) together with the limiting eigenvalues
(\ref{set-of-eigenvalues-1}) of the reduced spectral problem $LII$ (dash-dotted lines).
Notice that the results are presented in the context of the rescaled
variant of the Gross-Pitaevskii equation (\ref{GPphys})
commonly used in the physical literature, illustrating the relevant
eigenvalues as a function of the chemical potential $\mu = 1/(2 \eps)$.

The ground state $\eta_{\eps}$ exists for any $\eps < \frac{1}{2}$ (i.e., $\mu>1$)
and the limit $\eps \to \frac{1}{2}$ can be obtained via small-amplitude bifurcation theory \cite{zezyulin}.
All eigenvalues $\gamma = -\lambda^2$ in the spectral problem (\ref{LL}) 
in this limit occur at the integers $4 (n + m)^2$ with $n,m \geq 0$ and
the multiplicity of the eigenvalue $\gamma = 4(n + m)^2$ is $n+ m + 1$.
When $\eps < \frac{1}{2}$, this degeneracy is broken and all eigenvalues
become smaller as $\eps$ gets smaller (or $\mu$ increases)
besides the double eigenvalue $\gamma = 4$ and the simple
eigenvalue $\gamma = 16$. Notice that the eigenvalues $\gamma$
are related to the eigenfrequencies $\omega$ on Figure \ref{fig1} by 
$\omega=\sqrt{\gamma}/2$ (accounting for the time rescaling $\tau = 2 t$).

Persistence of $\eps$-independent eigenvalues $\gamma = 4$ and $\gamma = 16$ 
of the spectral problem (\ref{LL}) is explained by
the symmetries of the Gross--Pitaevskii equation (\ref{GP}). 
One symmetry is given by the explicit transformation of solutions
\begin{equation}
\label{symmetry-1}
u(x,y,t) = e^{i p(t) x + i s(t) y + i \omega(t)} \tilde{u}(\tilde{x},\tilde{y},t), \quad
\tilde{x} = x - q(t), \;\; \tilde{y} = y - k(t).
\end{equation}
If $\tilde{u}(\tilde{x},\tilde{y},t)$ is a solution of equation (\ref{GP}) rewritten
in tilded variables and $(p,s,q,k,\omega)$ satisfy
$$
\left\{ \begin{array}{l} \dot{q} = 2 \eps p, \quad \eps \dot{p} + 2 q = 0, \\ \dot{k} = 2 \eps s,
\quad \eps \dot{s} + 2 k = 0, \end{array} \right. \quad \omega = -\frac{1}{2} (q p + ks),
$$
then $u(x,y,t)$ is also a solution of equation (\ref{GP}). 
Therefore, both $q$ and $k$ satisfy the linear oscillator equations with
eigenvalue $\gamma = 4$, which gives the double degeneracy of eigenfrequency $\omega = 1$ in Figure~\ref{fig1}.

The other symmetry of the Gross--Pitaevskii equation (\ref{GP}) with $d = 2$
is given by the conformal transformation
\begin{equation}
\label{symmetry-2}
u(x,y,t) = a(t) \tilde{u}(\tilde{x},\tilde{y},\tilde{t}) e^{i c(t) x^2 + i c(t) y^2+ i \omega(t)} , \quad
\tilde{x} = a(t) x, \;\; \tilde{y} = a(t) y, \;\; \tilde{t} = b(t),
\end{equation}
where $(a,b,c,\omega)$ satisfy the first-order differential equations
$$
\dot{b} = a^2, \quad \eps \dot{\omega} = 1 - a^2, \quad \dot{a} + 4 \eps a c = 0, \quad
\eps \dot{c} + 4 c^2 \eps^2 + 1 = a^4.
$$
Excluding $c$ and denoting $a(t) = z^{-1}(t)$, we obtained the nonlinear oscillator equation for $z(t)$:
$$
\ddot{z} + 4 z ( 1 - z^{-4}) = 0.
$$
There is a unique critical point $z = 1$ and it is a center
with eigenvalue $\gamma = 16$, corresponding to the eigenfrequency
$\omega = 2$ in Figure \ref{fig1}.

The two symmetries (\ref{symmetry-1}) and (\ref{symmetry-2}) 
explain the $\eps$-independent eigenfrequencies $\omega$ on Figure \ref{fig1}. On the other hand,
the figure shows that all eigenvalues $\gamma$ approach to the limiting eigenvalues
(\ref{set-of-eigenvalues-1}) as $\eps \to 0$ (i.e., as $\mu \rightarrow \infty$). 
This output confirms the robustness of the asymptotic distributions
presented herein. 

It is worth noting that in all the cases shown on Figure \ref{fig1} 
the eigenvalues have been confirmed including also their multiplicities.
For instance the eigenfrequency associated with $\omega=2$ ($\gamma=16$)
is associated with two eigenvalues in the set (\ref{set-of-eigenvalues-1}), 
namely with $k=1$ and $m=0$, as well as with $k=0$ and $m=4$. One of these 
corresponds to the conformal symmetry (\ref{symmetry-2}), 
while the other one can be observed on Figure \ref{fig1} to
asymptote to $\omega=2$ in the large-$\mu$ limit, as expected.

\section{Conclusions and future directions}

In the present work, we offered a systematic approach towards
identifying the eigenfrequencies of oscillations of the perturbations 
around the ground state of a Bose-Einstein condensate in an arbitrary number of
dimensions (our calculations were given in 1-, 2- and 3-dimensions).
This spectrum is important because it corresponds to the
excitations that can be (and have been) experimentally observed
once the condensate is perturbed appropriately. 

Part of the rationale for attempting to understand the details
of this spectrum is that when fundamental nonlinear excitations
are additionally considered on top of the ground state, then the
spectrum contains both a ``ghost'' of the spectrum of the ground state
and the so-called negative energy modes that pertain to the
nonlinear excitation itself. Relevant examples of this sort can
be found both for the case of one-dimensional dark solitons (and
multi-solitons) as analyzed in \cite{heidelberg} and in the
case of two-dimensional vortices, as examined in \cite{heidelberg2}. 
It is then of particular interest to try to understand eigenfrequencies of 
excitations of these structures in the Thomas-Fermi limit, as well as those of
their three-dimensional generalizations bearing line- or
ring-vortices. Such studies would be especially interesting
for future works.

{\bf Acknowledgments}: PGK is partially supported
by NSF-DMS-0349023 (CAREER), NSF-DMS-0806762 and the Alexander-von-Humboldt
Foundation. DEP is supported by the NSERC grant.


\begin{thebibliography}{99}

\bibitem{anderson} M.H.J. Anderson, J.R. Ensher, M.R. Matthews, C.E. Wieman,
``Observation of Bose-Einstein condensation in a dilute atomic
vapor'', Science {\bf 269}, 198--201 (1995).

\bibitem{davis} K.B. Davis,  M.-O. Mewes, M.R. Andrews, N.J. van Druten,
D.S. Durfee, D.M. Kurn and W. Ketterle, ``Bose-Einstein condensation in a gas of sodium atoms'',
Phys. Rev. Lett. {\bf 75}, 3969--3973 (1995).

\bibitem{book1} C.J. Pethick and H. Smith,
{\it Bose-Einstein condensation in dilute gases}, Cambridge University
Press (Cambridge, 2002).

\bibitem{book2}  L.P. Pitaevskii and S. Stringari,
{\it Bose-Einstein Condensation}, Oxford University Press (Oxford, 2003).

\bibitem{review} F. Dalfovo, S. Giorgini, L.P. Pitaevskii and S. Stringari,
``Theory of Bose-Einstein condensation in trapped gases'',
Rev. Mod. Phys. {\bf 71}, 463--512 (1999).

\bibitem{rcg:BEC_BOOK}
P.G. Kevrekidis, D.J. Frantzeskakis, and  R. Carretero-Gonz{\'a}lez (eds.).
{\sl Emergent Nonlinear Phenomena in Bose-Einstein Condensates:
Theory and Experiment}. Springer Series on Atomic, Optical, and Plasma Physics {\bf 45}
(Springer, Heidelberg, 2008).

\bibitem{rcg:65}
R. Carretero-Gonz{\'a}lez, D.J. Frantzeskakis, and P.G. Kevrekidis.
``Nonlinear Waves in Bose-Einstein Condensates:
Physical Relevance and Mathematical Techniques'', Nonlinearity
{\bf 21}, R139--R202 (2008).

\bibitem{sulem} C. Sulem and P.L. Sulem,
\newblock {\it The Nonlinear Schr{\"o}dinger Equation},
Springer-Verlag (New York, 1999).

\bibitem{ablowitz} M.J. Ablowitz, B. Prinari and A.D. Trubatch,
{\it Discrete and Continuous Nonlinear Schr{\"o}dinger Systems},
Cambridge University Press (Cambridge, 2004).

\bibitem{kevfra}  P.G. Kevrekidis and D.J. Frantzeskakis,
``Pattern forming dynamical instabilities
of Bose-Einstein condensates'',
Mod Phys. Lett. B {\bf 18}, 173-202 (2004).

\bibitem{stringari} S. Stringari, ``Collective excitations of a trapped
Bose--condensed gas'', Phys. Rev. Lett. {\bf 77}, 2360--2363 (1996)

\bibitem{FCSG} M. Fliesser, A. Csordas, P. Szepfalusy, and R. Graham,
``Hydrodynamic excitations of Bose condensates in anisotropic traps'',
Phys. Rev. A {\bf 56}, R2533--R2536 (1997)

\bibitem{EGD} C. Eberlein, S. Giovanazzi, and D.H.J. O'Dell,
``Exact solution of the Thomas--Fermi equation for a trapped
Bose--Einstein condensate with dipole--dipole interactions'',
Phys. Rev. A {\bf 71}, 033618 (2005)

\bibitem{GalPel1} C. Gallo and D. Pelinovsky,
``Eigenvalues of a nonlinear ground state in the
Thomas--Fermi approximation'', J. Math. Anal. Appl. {\bf 355},
495-–526 (2009)

\bibitem{ZAKP} D.A. Zezyulin, G.L. Alfimov, V.V. Konotop, and
V.M. P\'erez--Garc\'ia, ``Stability of excited states of a Bose--Einstein condensate
in an anharmonic trap'', Phys. Rev. A {\bf 78}, 013606 (2008)

\bibitem{zezyulin} D.A. Zezyulin, ``Stability of two-dimensional radial excited states of
a Bose--Einstein condensate in an anharmonic trap", Phys. Rev. A {\bf 79}, 033622 (2009)

\bibitem{PeliKev} D.E. Pelinovsky and P.G. Kevrekidis,
``Periodic oscillations of dark solitons in parabolic potentials'',
Cont. Math. {\bf 473}, 159-179 (2008)

\bibitem{GalPel2} C. Gallo and D. Pelinovsky,
``On the Thomas--Fermi ground state in a radially symmetric
parabolic trap'', arXiv:0911.3913 (2009)

\bibitem{Grad} I.S. Gradshteyn and I.M. Ryzhik,
{\em Table of integrals, series and products}, 6th edition,
(Academic Press, 2005)

\bibitem{heidelberg} G. Theocharis, A. Weller, J.P. Ronzheimer, C. Gross,
M.K. Oberthaler, P.G. Kevrekidis and D.J. Frantzeskakis,
``Multiple atomic dark solitons in cigar-shaped Bose-Einstein
condensates'', arXiv:0909.2122.

\bibitem{heidelberg2} S. Middelkamp, P.G. Kevrekidis, D.J. Frantzeskakis,
R. Carretero-Gonz{\'a}lez and P. Schmelcher,
``Anomalous modes and matter-wave vortices in the presence of
collisional inhomogeneities and finite temperature'',
arXiv:0911.3308.

\end{thebibliography}
\end{document}